\documentclass{iau} 
\usepackage{graphicx}

\title[VMC proper motions] 
{VMC proper motions of the \\Magellanic Bridge}

\author[Thomas Schmidt \& Maria-Rosa Cioni]   
{Thomas Schmidt$^1$, Maria-Rosa Cioni$^1$, Florian Niederhofer$^1$, Jonathan Diaz$^2$, Gal Matijevic$^1$
}

\affiliation{$^1$Leibniz-Institut f\"{u}r Astrophysik Potsdam \\ An der Sternwarte 16, D-14482 Potsdam, Germany \\ email: {\tt tschmidt@aip.de}, \\ $^2$ International Centre for Radio Astronomy Research, The University of Western Australia,\\ M468, 35 Stirling Hwy, Crawley WA 6009, Australia}

\pubyear{2018}
\volume{344}  
\jname{Dwarf Galaxies: From the Deep Universe to the Present}
\editors{S. Stierwalt \& K. McQuinn, eds.}
\begin{document}

\maketitle

\begin{abstract}
Dwarf galaxies enable us to study early phases of galaxy evolution and are key to many open questions about the hierarchical structure of the Universe. The Large and Small Magellanic Cloud (LMC and SMC) are the most luminous dwarf galaxy satellites of the Milky Way (MW). They are most likely gravitationally bound to each other, and their last interaction occurred about 200 Myr ago. Also, they are in an early phase of minor merging with the MW and will impact the Galactic structure in the future because of their relatively large mass. However, there are still major uncertainties regarding their origin and their interactions with one another and with the Milky Way. We cross-correlated the VMC and Gaia DR2 data to select a sample of stars that likely belong to the Magellanic Bridge, a feature formed of gas and stars which is connecting the LMC and the SMC. We removed potential MW foreground stars using a combination of parallax and colour-magnitude criteria and calculated the proper motions of the Bridge member stars. Our analysis supports a motion of star towards the LMC, which was found to be in good agreement with a dynamical simulation, of the SMC being stripped by the LMC.

\end{abstract}

\firstsection 
              
\section{Introduction}

The Magellanic Bridge, connecting the Magellanic Clouds, was first discovered by \cite[Hindman et al. (1963)]{Hindman63} as an over-density of neutral hydrogen gas. Subsequent studies have shown that the Bridge was likely formed tidally by stripping gas preferentially from the SMC during the last interaction with the LMC (\cite[Tsujimoto \& Bekki 2013]{TsujimotoBekki13}). Current estimates from \cite[Zivick et al. (2018)]{Zivick18}, based on stellar proper motion measurements, suggest this interaction to be quite recent (147$\pm$33 Myr ago).

The previously found young stellar populations of the Bridge by \cite[Irwin et al. (1985)]{Irwin85} should then have formed in situ. An older population of stars was expected to be present as well since tidal forces should have similar effects on stars and gas. Later observational studies (e.g., \cite[Bagheri, Cioni \& Napiwotzki 2013]{Bageri13}; \cite[N\"{o}el et al. 2013]{Noel13}) presented evidence of older populations in the Bridge, that were also supported by simulations (e.g. \cite[Guglielmo et al. 2014]{Gulielmo14}). The first spectroscopic evidence of a stellar population older than 1 Gyr between the Clouds was by \cite[Carrera et al. (2017)]{Carrera17}. The metallicity of this population suggests that it originated in the outer regions of the SMC. In contrast, stars formed from the stripped gas were shown to have metal abundances consistent with having formed in situ (\cite[Dufton et al. 2008]{Dufton08}). 

Measuring the proper motion of both young and old stars in the Magellanic Bridge can provide an important insight to further constrain dynamical simulations and answer open questions about the formation and evolution of the Magellanic Clouds.

\section{Observations}
Data analysed here are taken from the VISTA survey of the Magellanic Clouds system (VMC; \cite[Cioni et al. 2011]{Cioni11}). The VMC survey started acquiring data in November 2009, and it is almost complete. 
It consists of multi-epoch near-infrared observations in the $Y$, $J$, and $K_s$ band of 110 overlapping tiles across the Magellanic system: 68 covering the LMC, 27 covering the SMC, 13 across the Bridge and 2 within the Magellanic Stream. 
Each tile covers 1.77 deg$^2$ in the sky and results from a mosaic of 6 pawprints each containing 16 detectors. 
In this study, we focus on the 13 Bridge tiles. 
The VMC survey provides 12 $K_s$-band epochs as sensitive as $\sim19.2$ mag ($5\sigma$ Vega) and a spatial resolution of $<1^{\prime\prime}$“, which refers to $0.27$ pc at the mean distance of Bridge ($\sim55$ kpc). 

\section{VMC proper motion calculation}
The VMC proper motion of individual sources, contained within the 13 Bridge tiles, was calculated from a linear least square fit of the pixel displacements as a function of time with respect to a steady reference frame defined by background galaxies. Each fit contained on average 10 data points, with a minimum of 8, spanning an average baseline of 921 days. Calculations were done on a detector level separately for \textit{x} and \textit{y} directions for each of the 16 detectors and 6 pawprints of each tile. The median proper motion of background galaxies was calculated to validate the reference frame and was then subtracted from the stellar proper motions to account for systematic uncertainties. The total number of sources in the resulting proper motion catalogue are 320,246 stars and 468,122 background galaxies. This includes independently measured proper motions from duplicates caused by the overlap of adjacent VMC tiles. After the removal of MW foreground stars, we expect less than $10.4\%$ of these stars to be members of the Bridge.

\section{Milky Way foreground contamination}
A major issue to determine the proper motion of Magellanic Bridge stars is the influence of Milky Way foreground stars. To lower this contamination we used data from Gaia DR2 cross-matched with the VMC survey. As a first step, we removed stars with parallaxes $\omega>0.2$. This removes foreground stars up to $12$ kpc with a declining efficiency due to the nature of Gaia parallaxes (\cite[Luri et al. 2018]{Luri18}).

\begin{figure}[ht]
\begin{center}
 \includegraphics[width=5.1in]{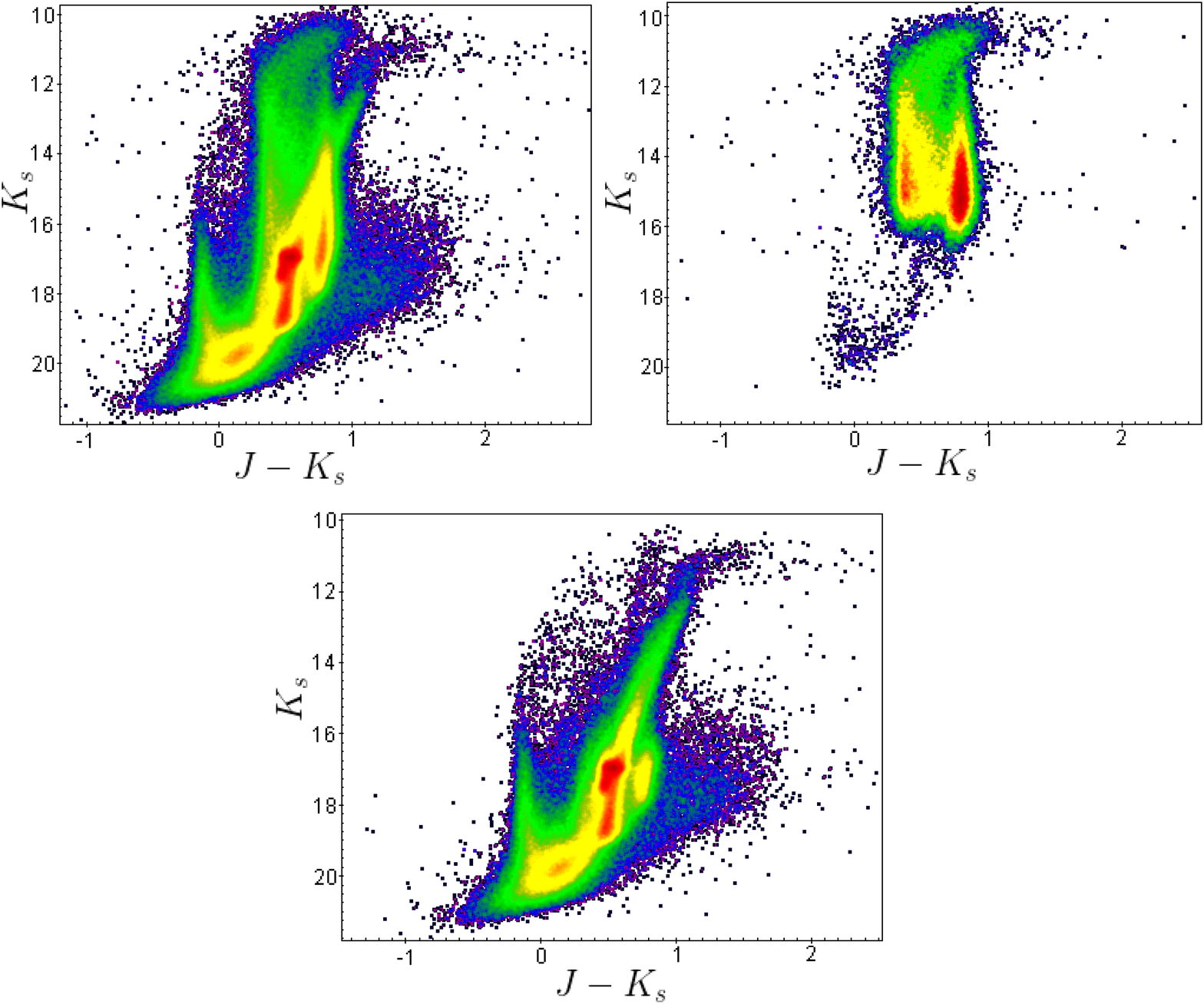} 
 \caption{Colour-magnitude diagrams of cross-matched stars of the Magellanic Bridge from VMC and Gaia DR2. The top left panel presents all cross-matched stars, the one top right shows a selection of Milky Way stars (with $\omega>0.2$) whereas the bottom panel shows the remaining stars after the removal of the foreground stars.}
   \label{figCMDs}
\end{center}
\end{figure}

To further reduce the influence of possible Milky Way stars, we also used selection criteria in a ($J-K_{s}$) vs $K_{s}$ colour-magnitude diagram (CMD), based on VMC data. Figure 1 shows three CMDs of a VMC-Gaia DR2 cross-matched sample before (left) and after (right) the removal of foreground stars (centre). Prominent features on these CMDs, related to the Magellanic Clouds and Bridge, are young main sequence stars (at $J-K_{s}=-0.2$ and $K_{s}=15-19$), red clump stars (at $J-K_{s}=0.5$ and $K_{s}\sim17$), red giant branch stars (at $J-K_{s}=0.7-1.0$ and $K_{s}=12-16$), red supergiants (at $J-K_{s}\sim0.6$ and $K_{s}=11-13$), and asymptotic giant branch stars (at $J-K_s=0.8-2.0$ and $K_s=11-12$). A large majority of these stars are found towards LMC and SMC, while the central regions show a low stellar density, this can be seen in underlying density distribution in Figure 2. The Milky Way foreground manifests itself as two long nearly vertical features, at $J-K_{s}\sim0.3$ caused by the main sequence turn-off of populations of intermediate to old ages, and at $J-K_{s}\sim0.8$ caused by low-mass cool M dwarfs.

\section{Proper motion of the Magellanic Bridge}
Due to the difficulty of measuring proper motions at the distance of the Bridge, proper motion errors are too large to separate stellar populations of the Bridge itself from those of the Milky Way foreground, the LMC and SMC in proper motion space. Furthermore, the Bridge stellar population density is much lower than that of the Milky Way foreground. 

To determine the median proper motion of the Bridge stars we then proceeded as follows. We used two-dimensional Voronoi binning to divide the Bridge into 40 spatial bins with each bin holding a minimum of 50 stars. Then, we determined the median proper motion of the stars within each bin. We performed these steps separately for proper motions derived from the VMC data and from the Gaia DR2.
Both proper motions values were compared to proper motions derived from a dynamical N-body simulation of the SMC stripped under the influence of the LMC mass (\cite[Diaz \& Bekki 2012]{DiazBekki12}). Figure 2 presents the results of this analysis. It shows the proper motion map of Magellanic Bridge derived from Gaia DR2 proper motions compared with the simulation. Proper motion uncertainties amount to $\sim 0.5$ mas yr$^{-1}$.

\begin{figure}[ht]
\begin{center}
 \includegraphics[width=5.1in]{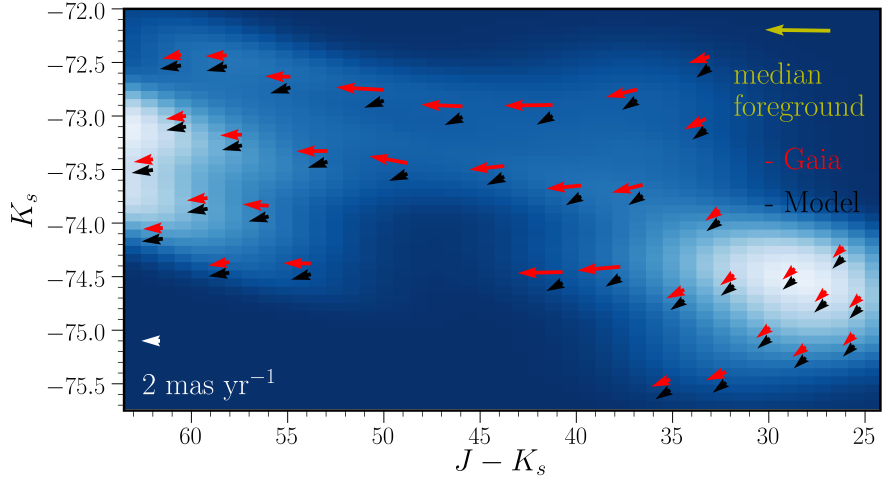}
 \caption{Proper motion map of the Magellanic Bridge using Gaia DR2 measurements (red) compared to a dynamical simulation from \cite[Diaz \& Bekki (2012)]{DiazBekki12} (black) superimposed to the stellar density of the cross-matched VMC and Gaia samples. The median proper motion of the Milky Way foreground is indicated in the top left}
   \label{figPM_map}
\end{center}
\end{figure}

\section{Conclusion}
The first proper motion measurements of stars across the Magellanic Bridge from VMC and Gaia DR2 indicate that stars move from the SMC to the LMC (see Fig.\,\ref{figPM_map}). This supports simulations of the stripping of the SMC resulting from the dynamical interaction with the LMC. Assuming this scenario is correct and that uncertainties in the proper motion measurements are properly estimated, remaining discrepancies between model and observations may be attributed to a residual Milky Way foreground contamination. 

This study proves that a combination of Gaia DR2 and VMC data can select probable Magellanic bridge stars across a wide area. In the future, we plan to extend the near-infrared observations to other tiles in the central regions of the Bridge and also to acquire additional VISTA epochs to improve the measurement of proper motions. Gaia parallaxes, that we used to subtract the Milky Way foreground stars, are expected to improve in the next data releases.


\end{document}